# NEW XOR & XNOR OPERATIONS IN INSTANTANEOUS NOISE-BASED LOGIC

NASIR KENARANGUI [§], ARTHUR POWALKA, LASZLO B. KISH

*Department of Electrical and Computer Engineering, Texas A&M University, College Station, TX 77843-3128, USA*

Instantaneous Noise-Based Logic (INBL) is a classical noise-based computing framework that serves as an alternative to quantum computation. However, some of the logic gates remain unimplemented for achieving gate universality over superpositions. INBL encodes $M$ noise-bits using $2M$ orthogonal stochastic reference noises to construct a $2^M$-dimensional product-based Hilbert space (hyperspace). Each vector in this hyperspace is the product of $M$ noise-bits, and each noise-bit corresponds to one of two reference noises. This work introduces INBL implementations of XOR and XNOR gate operations formulated for the symmetric INBL scheme. The resulting gate algorithms support pairwise operations between strings of arbitrary length and naturally extend to superpositions. This offers the potential for substantial gains in computational speed and reductions in hardware complexity. In particular, the XOR and XNOR gates can act on inputs drawn from any combination of individual noise-bits, full bit strings, or superpositions. Diverging from previous methods by Khreishah et al., our approach avoids direct manipulation of the reference noise system, enabling more flexible and general-purpose implementations. We validate NOT, XOR, and XNOR operations in symmetric INBL, through simulations using random telegraph waves (RTW), demonstrating practical feasibility without explicit reference noise manipulation.

## 1. Introduction

In this paper, we introduce a new solution for the XOR and XNOR operations in instantaneous noise-based logic (INBL). For the sake of the reader, we first do a brief review.

---

[§] Corresponding Author.

### 1.1. *Instantaneous noise-based logic*

In noise-based logic (NBL), the logic information is represented by noise-bits constructed from the relevant reference signals [1,2]. Logic operations can be executed on these reference signals. The reference noise system (RNS) is based on a truly random number generator (TRNG); see Figure 1.

INBL [3-15] is a subclass of NBL that does not require time averaging (as a part of correlators) to distinguish between logic states. This allows the results of the operations to be obtained "instantaneously". (The name INBL is from Ferdinand Peper [5] at Institute of Communications Technology, Japan).

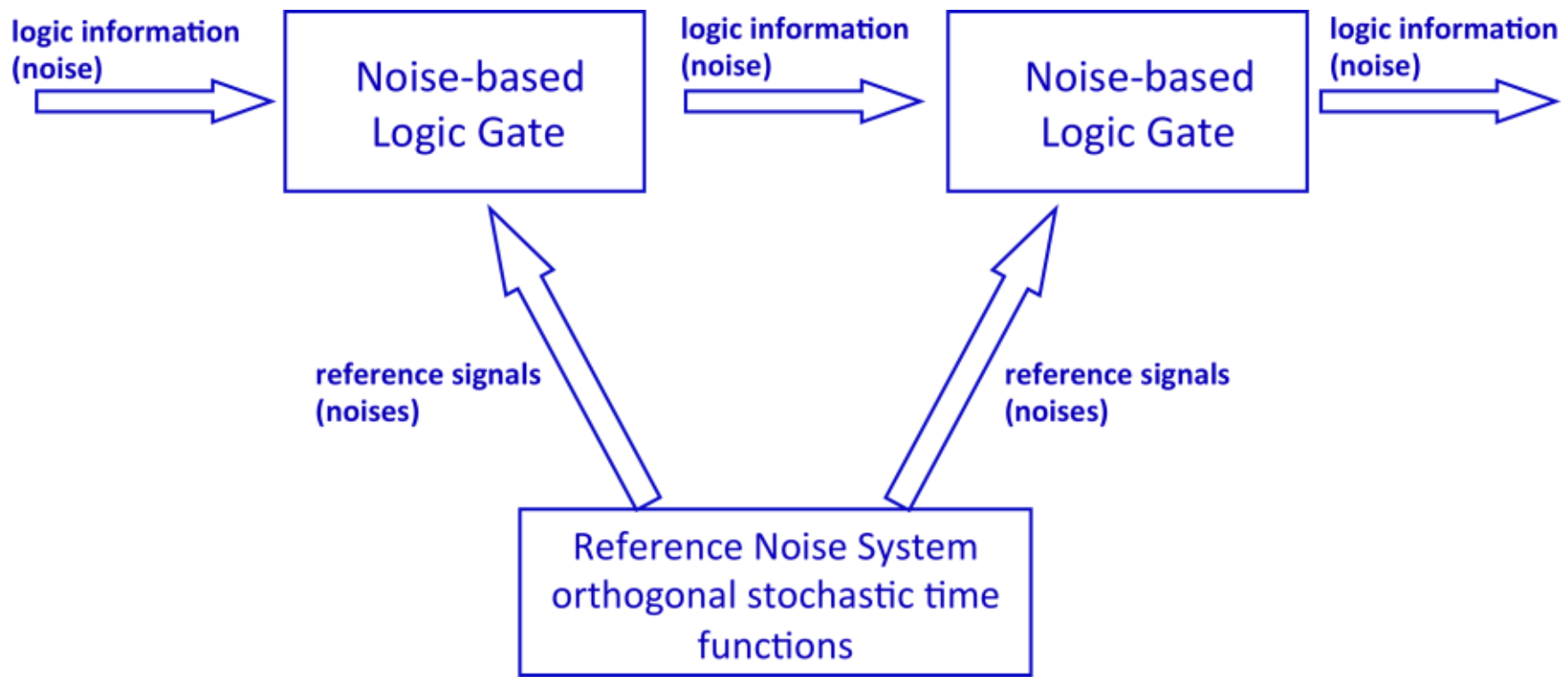

**Figure 1.** Generic noise-based logic hardware scheme [1,3,4]. Logic operations can be executed by the gates or by operations on the reference signals. The reference noise system is based on a TRNG. The interesting INBL schemes are mostly those that can handle exponentially large data sets and can be implemented by a binary classical computer (Turing machine) with a bit resolution and complexity that is polynomial in $M$., e.g. [3,4,7,9-15].

Each logic state is defined by a combination of reference noises which are independent stochastic processes. $M$ noise-bits require $2M$ uncorrelated (orthogonal) reference noises [1-7], see Figure 2. Each noise-bit $i$ is associated with a logical low reference noise $L_i(t)$ and a logical high reference noise $H_i(t)$, where $i = 1, \dots, M$. These $2M$ uncorrelated reference noises are represented by random telegraph waves (RTWs). These RTWs are independent random square waves with values of either $+1$ or $-1$ with a probability of 0.5 at the beginning of each clock cycle; see an example in Figure 3.

An $M$ bit binary product string, composed of reference noises, is called a hyperspace vector [1]. It corresponds to a decimal number $n$, where $n = 0,1, \dots, 2^{M-1}$ for an $M$-bit INBL system.

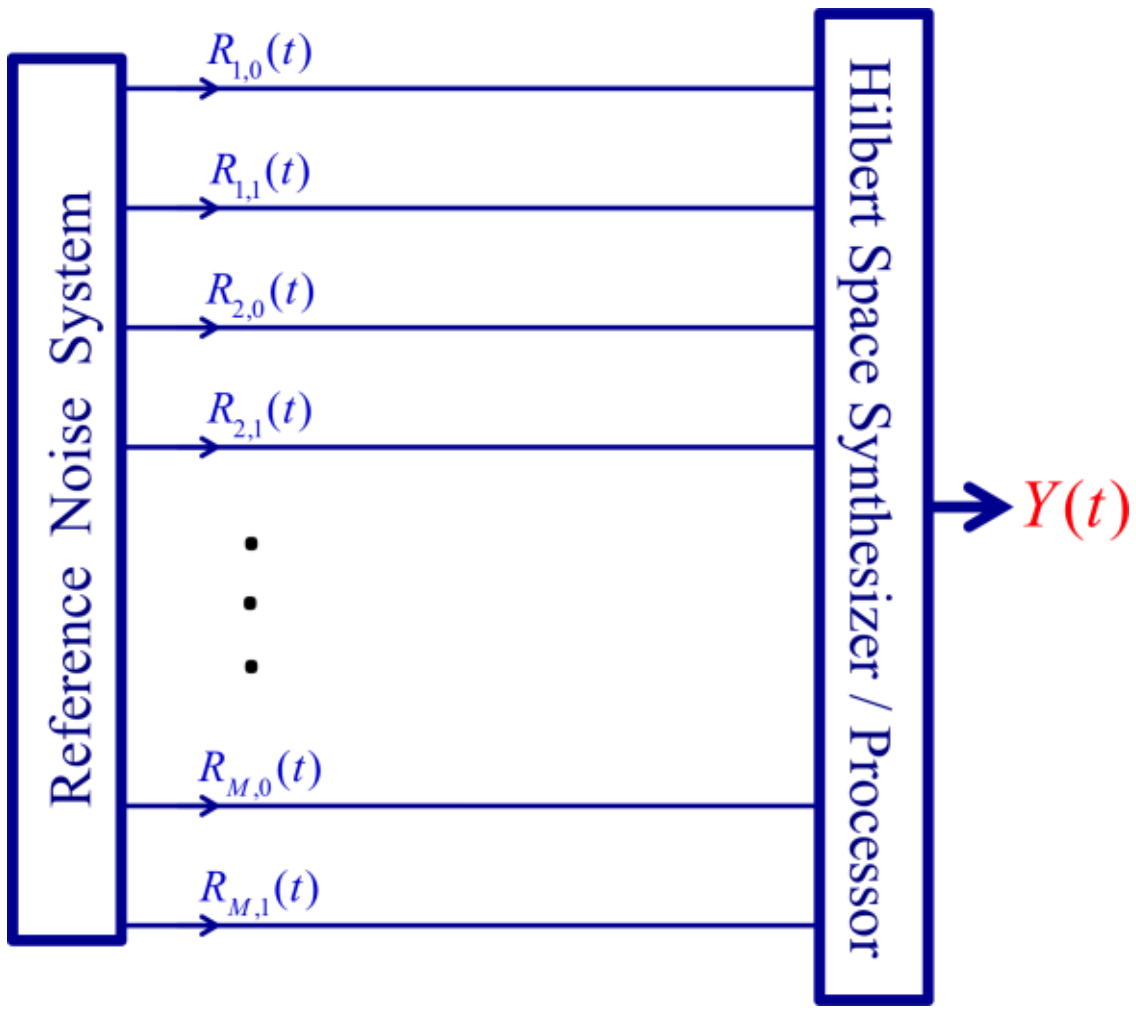

**Figure 2.** General example of an INBL system [3] with a Hilbert space synthesizer. This circuit illustration shows the structure of the generic superposition synthesizer. The signal $Y(t)$ represents a fraction or the whole Hilbert space with $2^M$ dimensions.

Each $n$ is thus represented by the reference noises $\{G_1^n(t), G_2^n(t), \ldots, G_M^n(t)\}$ [3-7]. The corresponding hyperspace vector, $S_n(t)$ of the $2^M$ dimensional Hilbert space, is generated by the bit string products of the relevant reference noises:

$$S_n(t) = \prod_{i=0}^{M-1} G_i^n(t). \tag{1}$$

For each bit $i$, $G_i^n(t) = L_i(t)$ when the bit value is a logic low, and $G_i^n(t) = H_i(t)$ when the bit value is a logic high. In a symmetric scheme, the RTWs for all of the reference noises, representing both $L_i(t)$ and $H_i(t)$, have the same amplitude range. For special purpose applications, product strings shorter than $M$ noise-bits are also possible, but a full hyperspace vector is $M$ bit long.

The superposition of all hyperspace vectors is called the universe, which can be generated with polynomial complexity using the Achilles' Heel algorithm [2,3]:

$$U(t) = \sum_{k=0}^{2^{M-1}} S_n(t) = \prod_{i=0}^{M-1} [L_i(t) + H_i(t)]. \tag{2}$$

This is similar to how in quantum computing the Hadamard operation is used to generate superpositions.

### 1.2. *The INBL Engine*

The INBL engine, developed for the purposes of simulation and analysis, is a software tool used to generate and manipulate orthogonal noise-bit signals. As an example, Figure 3 shows the reference noises for the first noise bit: $R_{10}(t)$ and $R_{11}(t)$. In the symmetric RTW scheme, all reference noises have the same discrete amplitude range of $\pm 1$ [7]. The INBL engine uses a TRNG to generate the RTW reference noises (logic low and logic high) for each of the $M$ noise-bits [1-6], resulting in $2M$ orthogonal RTWs for the reference noises:

$$R_{10}(t), R_{11}(t), R_{20}(t), R_{21}(t), \dots, R_{M0}(t), R_{M1}(t). \tag{3}$$

Note, that the reference noises are kept in computer memory [12,13]. The minimum number of RTW clock cycles required for the current INBL operations is in the order of 100, which is based on the minimum theoretical error probability of an idealistic gate [16].

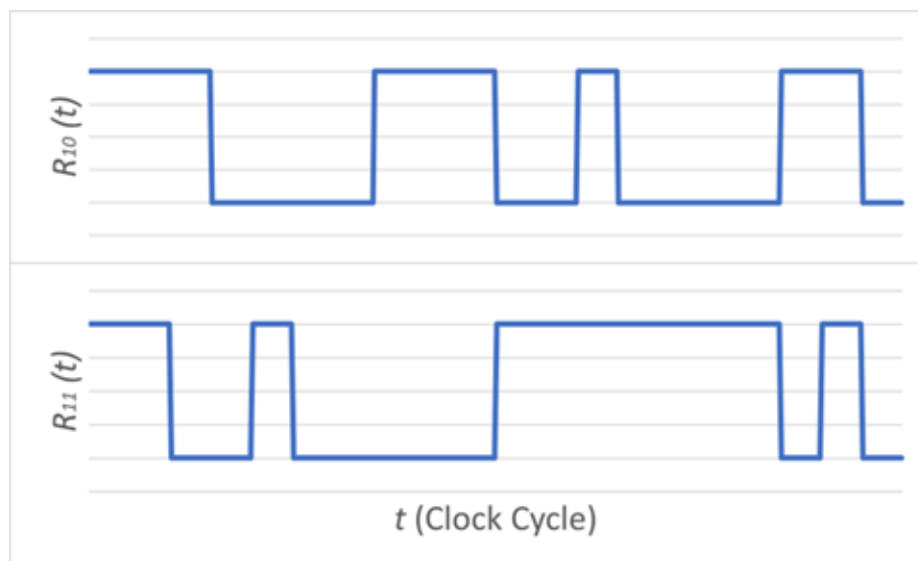

**Figure 3.** The first 20 clock cycles in $M = 4$ bit INBL system, where the symmetric RTWs (reference noises) for the first noise-bit are $R_{10}(t)$ and $R_{11}(t)$.

Hyperspace vectors are themselves RTWs constructed through the products of the relevant references noises as shown in Equation 1. Hyperspace vectors $\{A(t), B(t), C(t) \dots\}$ can be constructed to represent an arbitrary set of numbers $\{\mathrm{A}, \mathrm{B}, \mathrm{C}, \dots\}$:

$$\begin{aligned} A(t) &= G_1^{\mathrm{A}}(t)\, G_2^{\mathrm{A}}(t)\, G_3^{\mathrm{A}}(t) \dots G_M^{\mathrm{A}}(t) \\ B(t) &= G_1^{\mathrm{B}}(t)\, G_2^{\mathrm{B}}(t)\, G_3^{\mathrm{B}}(t) \dots G_M^{\mathrm{B}}(t) \\ C(t) &= G_1^{\mathrm{C}}(t)\, G_2^{\mathrm{C}}(t)\, G_3^{\mathrm{C}}(t) \dots G_M^{\mathrm{C}}(t) \\ &\vdots \end{aligned} \tag{4}$$

Consider for example, the binary number $\mathrm{A} = 0011$ represented in an $M = 4$ INBL system. It is represented by the hyperspace vector $A(t) = R_{10}(t)R_{20}(t)R_{31}(t)R_{41}(t)$. An arbitrary number of hyperspace vectors can be superimposed [10]:

$$Y(t) = A(t) + B(t) + C(t) + \cdots \tag{5}$$

Such superpositions enable transmission of large sets of numbers on a single wire without signal multiplexing, as well as parallel operations on these sets. As an example, consider the two additional binary numbers: $\mathrm{B} = 0101$ and $\mathrm{C} = 0001$. Figure 4 demonstrates the superposition of these numbers; that is, the sum $A(t) + B(t) + C(t)$ of the corresponding hyperspace vectors.

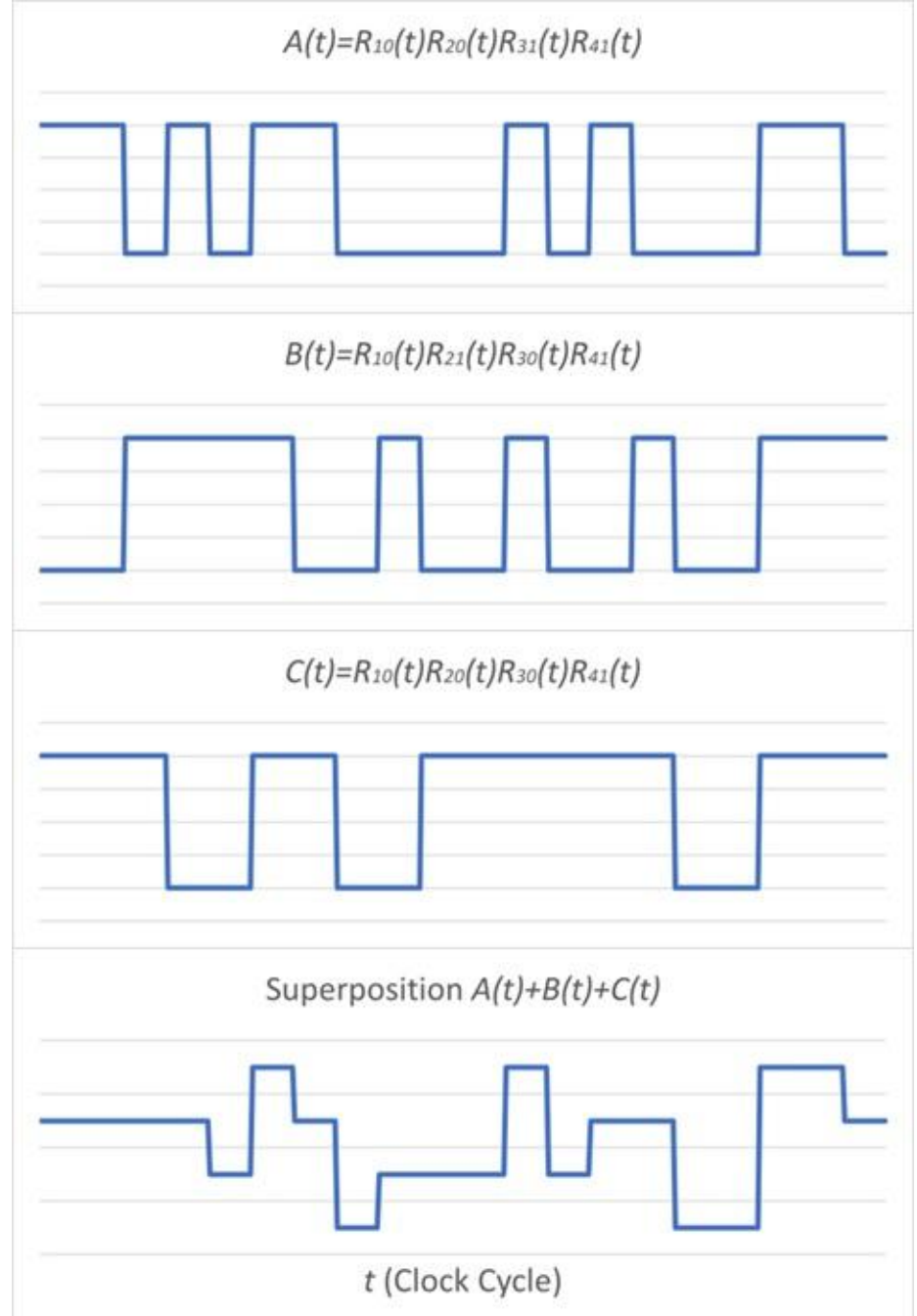

**Figure 4.** The first 20 clock cycles in $M = 4$ bit INBL system, where the hyperspace vectors $A(t)$, $B(t)$, $C(t)$ and their superposition are generated using the INBL engine. Note, the scale of the amplitude axis of the superposition is suppressed.

### 1.3. *Survey of the NOT INBL Operation*

As a steppingstone, this section surveys the NOT operation in INBL. This naturally leads to the introduction of the XOR and XNOR operations described in Section 2. Since multiplication is distributive over addition, any product operation applied to a superposition of hyperspace vectors extends directly to each hyperspace vector in the superposition. This feature enables inherently parallel computation executed simultaneously on all constituent states within the superposition [2-15].

The NOT operation (e.g. [6]) was developed for a reference system using symmetric RTWs (that is, with amplitudes $\pm 1$). The NOT operation on the $i$-th noise-bit is defined as:

$$\mathrm{NOT}_i(t) = R_{i0}(t)R_{i1}(t). \tag{6}$$

Take for example:

$$\mathrm{NOT}_1(t)R_{10}(t) = [R_{10}(t)R_{11}(t)]R_{10}(t) = R_{11}(t) \tag{7}$$

and

$$\mathrm{NOT}_1(t)R_{11}(t) = [R_{10}(t)R_{11}(t)]R_{11}(t) = R_{10}(t). \tag{8}$$

This operation targets a specific bit $i$ within a (product) string. Due to the commutativity of multiplication, these relations also work on strings, that is, on hyperspace vectors. For example, the operation $\mathrm{NOT}_1(t) = R_{10}(t)R_{11}(t)$ targets the LSB (Least Significant Bit) component $i = 1$.

Consider the hyperspace vector $A(t)$ for the 4-bit number $\mathrm{A} = 0011$. The LSB on string A can be flipped to get $\mathrm{A}_1^* = 0010$ using the $\mathrm{NOT}_1(t)$ operation on $A(t)$, resulting in the hyperspace vector for $A_1^*(t)$; see Figure 5.

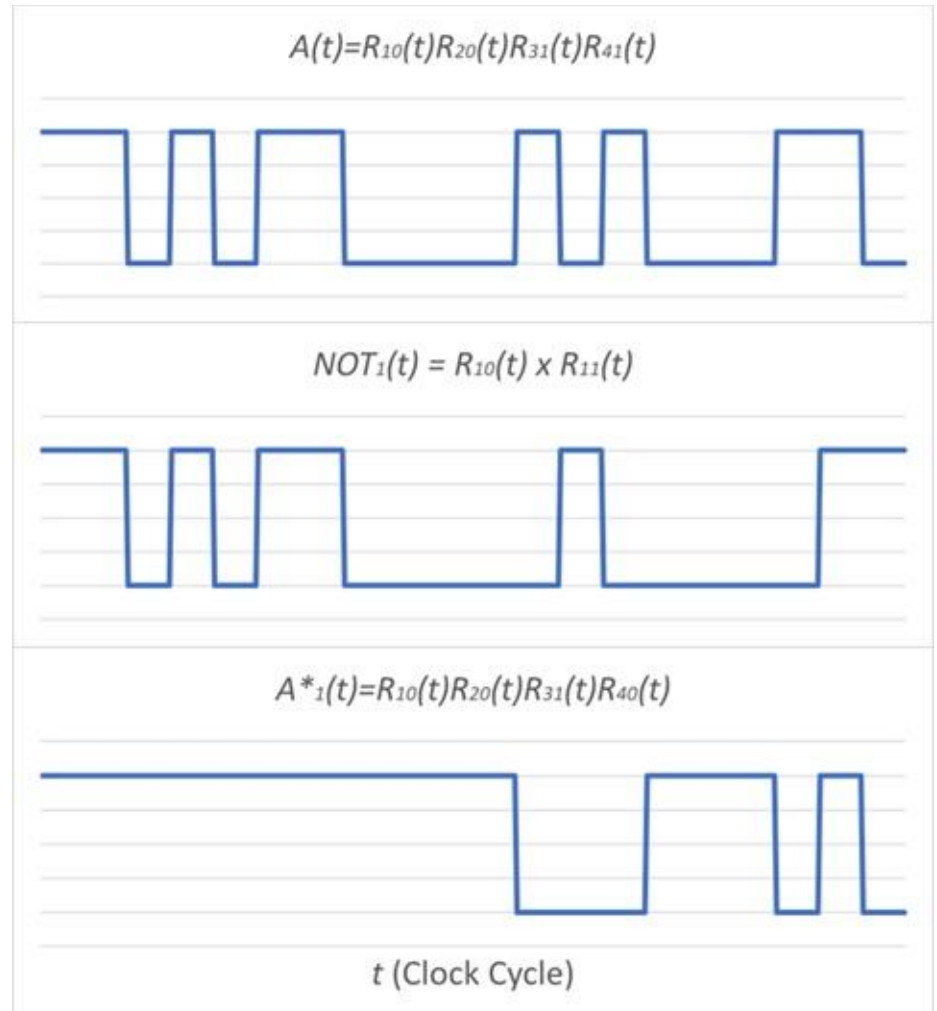

**Figure 5.** The first 20 clock cycles in an $M = 4$ bit INBL system; in this example, the $\mathrm{NOT}_1(t)$ operation on the hyperspace vector $A(t)$ is equivalent to inverting the LSB of the bit string A. $\mathrm{NOT}_1(t)$ is constructed from the reference $\mathrm{R}_{10}(t)$ and $\mathrm{R}_{11}(t)$ RTWs shown in Figure 3.

NOT operations can also be constructed to simultaneously target multiple bits. For example, if we want to simultaneously flip two noise bits $i = 1$ and $i = 3$, the NOT operation becomes:

$$\mathrm{NOT}_{13}(t) = R_{10}(t)R_{11}(t)R_{30}(t)R_{31}(t). \tag{9}$$

In INBL sets of numbers are represented by superpositions of hyperspace vectors. This algorithm for the NOT operation can be executed in parallel on these superpositions. Consider the superposition of hyperspace vectors $A(t), B(t)$, and $C(t)$ associated with $\{0101, 0001, 0011\}$; see Figure 2. Implementing $\mathrm{NOT}_{13}(t)$ on this superposition results in the superposition of $A_{13}^*(t), B_{13}^*(t)$, and $C_{13}^*(t)$ representing set $\{0000, 0100, 0110\}$; see Figure 6.

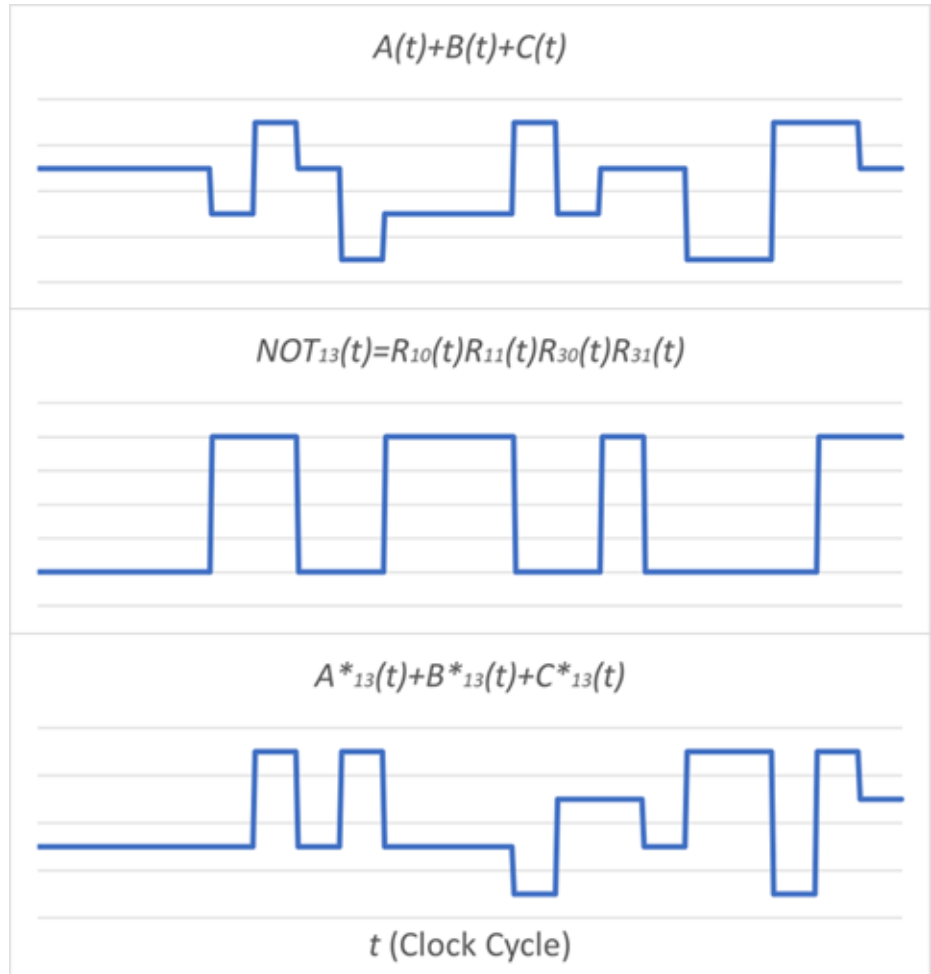

**Figure 6.** The first 20 clock cycles in an $M = 4$ bit INBL system. This example shows the $\mathrm{NOT}_{13}(t)$ operation targeting bits $i = 1$ and $i = 3$ on the superposition of hyperspace vectors $A(t), B(t)$, and $C(t)$ which represent the set of numbers $\{0101, 0001, 0011\}$. The result is the superposition of hyperspace vectors $A_{13}^{*}(t), B_{13}^{*}(t)$, and $C_{13}^{*}(t)$ representing the set of numbers $\{0000, 0100, 0110\}$. Note, the scale of the amplitude axis of the superpositions is suppressed.

This highlights a powerful feature of INBL: simultaneous operations on bits can be carried out on superpositions at a constant time complexity regardless of how many numbers are included in the superposition. This is due to the distributive property of multiplication, where the NOT operation on a hyperspace vector can also be carried out on a superposition of an arbitrary number of hyperspace vectors.

## 2. XOR and XNOR operations in the INBL system

A previous work by Khreishah, et al, [14] introduced bitwise XOR and XNOR operations based on manipulation of the reference wires, which builds on a concept introduced in [9,10]. However, the manipulation of reference wires can destroy the logic relations in other operations; thus, it must be used very cautiously [11]. The Khreishah XOR and XNOR gates had inputs of bit values in two separate noise-bits in a hyperspace vector, and the output was obtained as the value of a third noise bit in the same string.

Below we introduce a new algorithm for XOR and XNOR operations in INBL. It does not require manipulation or "grounding" of reference signals like in [3,9-11]. Additionally, the algorithm introduced below is not just an XOR and XNOR operation on individual bits but can also be a pairwise operation between hyperspace vectors, or between a hyperspace vector and an arbitrary large superposition. However, the input strings must have the same number of noise bits; this is automatically satisfied with hyperspace vectors.

As a starting point, we define the XOR operation for individual noise-bits $G_i^j(t)$ and $G_i^k(t)$, where $j, k \in \{0, 1\}$. When $j \neq k$, the output of the XOR operation must be 1, and when $j = k$, the output of the XOR operation must be 0. The operational property of the product of two noise-bits is as follows:

$$G_i^j(t)G_i^k(t) = \begin{cases} 1, & if\ j = k \\ \mathrm{NOT_i}(t), & if\ \ j \neq k \end{cases} \quad (10)$$

Consequently, the output of $\mathrm{XOR_i}(t)$ is achieved with simple multiplication of the above expression with the low noise-bit value $G_i^0(t)$:

$$\mathrm{XOR_i}(t)[G_i^j(t)G_i^k(t)] = G_i^j(t)G_i^k(t)G_i^0(t) = \begin{cases} G_i^0(t), & if\ j = k \\ \mathrm{NOT_i}(t)G_i^0(t) = G_i^1(t), & if\ \ j \neq k \end{cases} \quad (11)$$

Conversely, when $j \neq k$, the output of the XNOR operation must be 0, and when $j = k$, the output of the XNOR operation must be 1. This is achieved through multiplication with the one-noise-bit $G_i^1(t)$ instead:

$$\mathrm{XNOR_i}(t)[G_i^j(t)G_i^k(t)] = G_i^j(t)G_i^k(t)G_i^1(t) = \begin{cases} G_i^1(t), & if\ j = k \\ \mathrm{NOT_i}(t)G_i^1(t) = G_i^0(t), & if\ \ j \neq k \end{cases} \quad (12)$$

Obviously, these logic gates work even if one of the inputs is in a superposition of bit values. Then the output will also be in a superposition.

It is important to note that the pairwise XOR and XNOR algorithms leverage the fact that when the reference noise-bits are equal $(j = k)$, their product forms a constant vector of ones, known as the "vacuum state" which was introduced in the ternary INBL scheme [15]. In contrast, when the noise-bits differ $(j \neq k)$, the product of the reference noises creates the inverter described in the previous section. This unique property has a significant implication for operations in the INBL system: when two noise-bit values are identical $(j = k)$, these algorithms cannot distinguish between the cases $(j = k = 0)$ and $(j = k = 1)$ in accordance with the XOR and XNOR rules. This limitation poses a challenge for developing AND and OR operations using simple product-based methods.

These gates can be expanded for pairwise operations between bit strings. Consider strings A and B with the same length. If the strings are at maximum bit resolution, that is, they are hyperspace vectors of the $M$ noise-bit system, then:

$$A(t) = G_1^j(t)G_2^k(t)\ \dots\ G_M^l(t) \quad and \quad B(t) = G_1^m(t)G_2^n(t)\ \dots\ G_M^p(t). \quad (13)$$

The hyperspace vector of a bit string of zeros is defined as:

$$Zeros(t) = G_1^0(t)G_2^0(t)\ \dots\ G_M^0(t). \quad (14)$$

Using the commutative property, the pairwise XOR operation between bit strings A and B can be defined as:

$$A(t)B(t)Zeros(t) = [G_1^j(t)G_1^m(t)G_1^0(t)]\ [G_2^k(t)G_2^n(t)G_2^0(t)]\ \dots \\ \dots [G_M^l(t)G_M^p(t)G_M^0(t)] \quad (15)$$

The hyperspace vector of a bit string of all ones is defined as:

$$Ones(t) \;=\; G_1^1(t)G_2^1(t) \;\ldots\; G_M^1(t) \tag{16}$$

The pairwise XNOR operation between $A(t)$ and $B(t)$ then becomes:

$$A(t)B(t)Ones(t) \;=\; [G_1^j(t)G_1^m(t)G_1^1(t)]\,[G_2^k(t)G_2^n(t)G_2^1(t)] \;\ldots \\ \ldots [G_M^l(t)G_M^p(t)G_M^1(t)] \tag{17}$$

Consider the example of two $M = 4$ bit long strings $\mathrm{A} = 0011$ and $\mathrm{B} = 0101$, with the hyperspace vectors $A(t)$ and $B(t)$. The signals of $\mathrm{A\,XOR\,B} = 0110$ and $\mathrm{A\,XNOR\,B} = 1001$ are the expected outputs of the pairwise XOR and XNOR operations. Figures 7 and 8 demonstrate the functionality of these pairwise operations.

Much like the NOT operation, the pairwise XOR and XNOR operations can also be modified to target specific bits, while leaving the rest unchanged. For example, if we are targeting only the third bit of $A(t) = G_1^j(t)G_2^k(t)G_3^l(t)G_4^m(t)$, by XOR-ing it with the third noise-bit $G_3^p(t)$, the operation simplifies to:

$$A(t)\;\mathrm{XOR}_3\;G_3^p(t) = \;G_1^j(t)\;G_2^k(t)[G_3^l(t)G_3^p(t)G_3^0(t)]G_4^m(t), \tag{18}$$

When targeting the third noise-bit by XNOR-ing with $G_3^p(t)$, the operation simplifies to:

$$A(t)\;\mathrm{XNOR}_3\;G_3^p(t) = \;G_1^j(t)\;G_2^k(t)[G_3^l(t)G_3^p(t)G_3^1(t)]G_4^m(t). \tag{19}$$

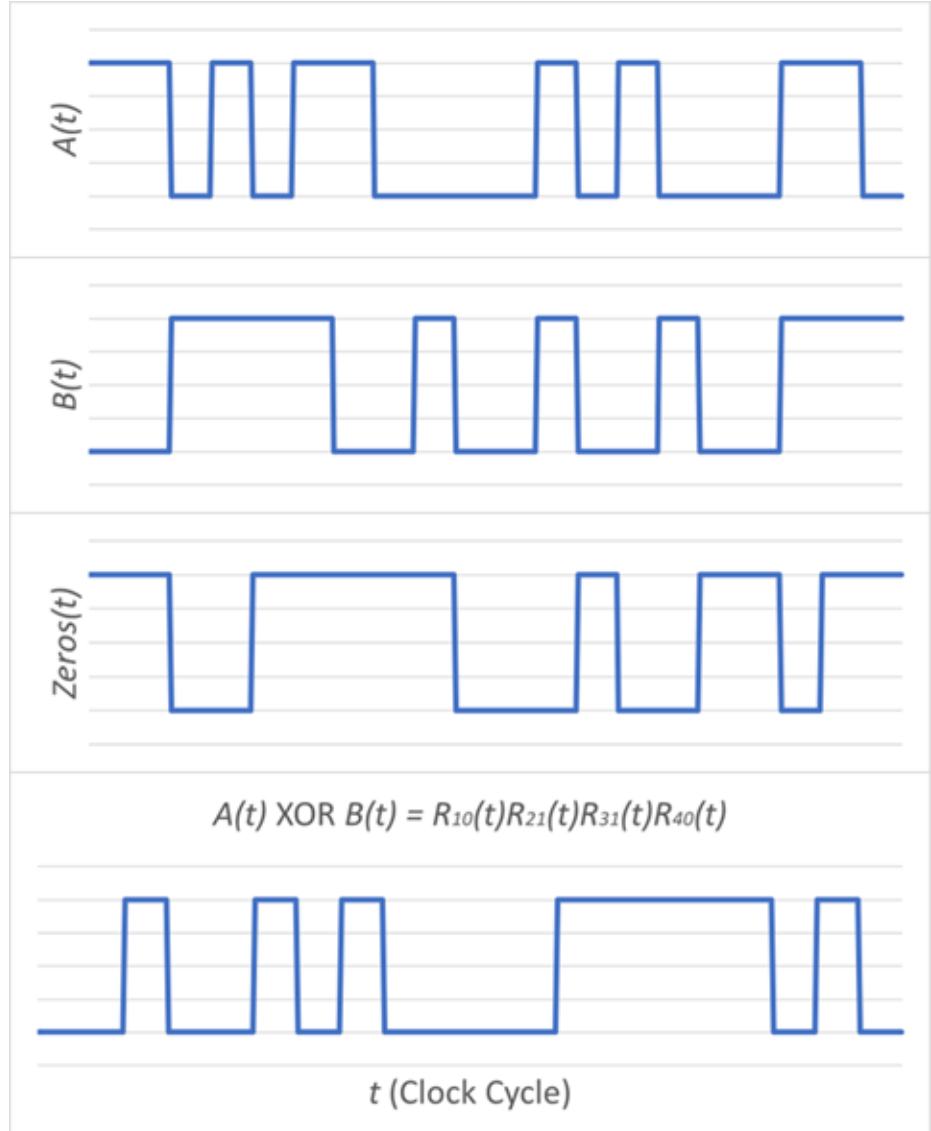

**Figure 7.** The first 20 clock cycles in an $M = 4$ bit INBL system. Shown above is the implementation of a pairwise XOR operation, where $\mathrm{A} = 0011$ and $\mathrm{B} = 0101$, with their respective hyperspace vector representations $A(t)$ and $B(t)$. The output is confirmed to be the hyperspace vector representation of bit string $\mathrm{A\,XOR\,B} = 0110$.

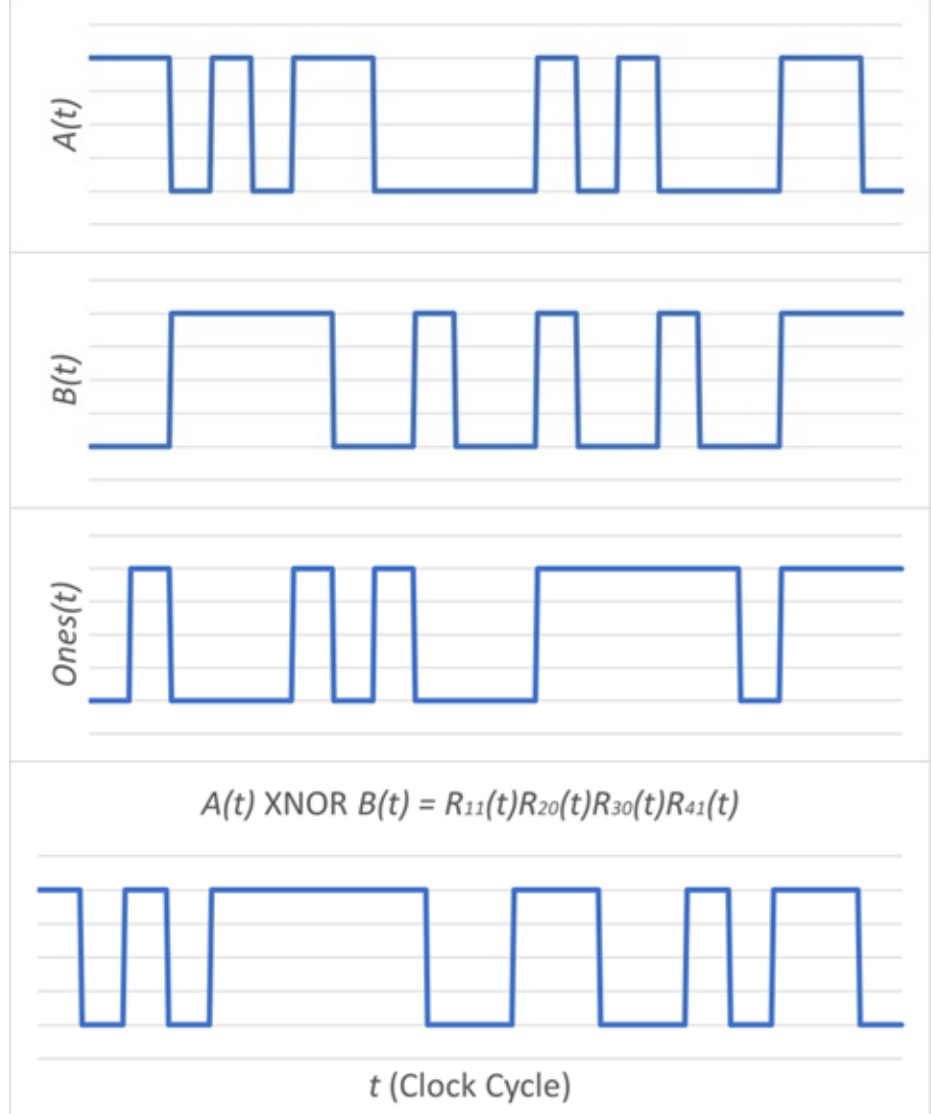

**Figure 8.** The first 20 clock cycles in an $M = 4$ bit INBL system. Shown above is the implementation of a pairwise XNOR operation, where $\mathrm{A} = 0011$ and $\mathrm{B} = 0101$, with their respective hyperspace vector representations $A(t)$ and $B(t)$. The output is confirmed to be the hyperspace vector representation of bit string A XNOR B $= 1001$.

The algorithms for the XOR and XNOR operations can be further expanded for superpositions, following a similar approach to the NOT operation explained in the previous section; see Figure 9 and 10.

| NOT With a Hyperspace Vector | | NOT With a Superposition | |
|---|---|---|---|
| **Input** | **Output** | **Input** | **Output** |
| Hyperspace Vector | Hyperspace Vector | Superposition | Superposition |
| $A(t)$ | NOT(t) $A(t)$ | $A(t)$ | NOT(t) $A(t)$ |
| | | $B(t)$ | NOT(t) $B(t)$ |
| | | $C(t)$ | NOT(t) $C(t)$ |
| | | ⋮ | ⋮ |
| **Pairwise XOR With a Hyperspace Vector** | | **Pairwise XOR With a Superposition** | |
| **Inputs** | **Output** | **Inputs** | **Output** |
| Hyperspace vector | Hyperspace Vector | Hyperspace vector | Superposition |
| $X(t)$ | $X(t)$ XOR $A(t)$ | $X(t)$ | $X(t)$ XOR $A(t)$ |
| Hyperspace vector | | Superposition | $X(t)$ XOR $B(t)$ |
| $A(t)$ | | $A(t)$ | $X(t)$ XOR $C(t)$ |
| | | $B(t)$ | ⋮ |
| | | $C(t)$ | |
| | | ⋮ | |
| **Pairwise XNOR With a Hyperspace Vector** | | **Pairwise XNOR With a Superposition** | |
| **Inputs** | **Output** | **Inputs** | **Output** |
| Hyperspace vector | Hyperspace Vector | Hyperspace vector | Superposition |
| $X(t)$ | $X(t)$ XNOR $A(t)$ | $X(t)$ | $X(t)$ XNOR $A(t)$ |
| Hyperspace vector | | Superposition | $X(t)$ XNOR $B(t)$ |
| $A(t)$ | | $A(t)$ | $X(t)$ XNOR $C(t)$ |
| | | $B(t)$ | ⋮ |
| | | $C(t)$ | |
| | | ⋮ | |

**Table 1.** The summary of gate operations for INBL. The NOT gate is a unary operation, while XOR and XNOR gates require two inputs and act pairwise between any combinations of noise-bits, hyperspace vectors, or superpositions (all combinations not shown in this table).

The NOT, XOR, and XNOR operations apply to sets of bit strings (represented by superimposed hyperspace vectors). However, there is one notable difference; while the NOT operation inverts individual bits within a string, XOR and XNOR are pairwise operations requiring two inputs. The distributive property of multiplication guarantees that pairwise XOR and XNOR operations between hyperspace vectors can distribute over a sum of hyperspace vectors, that is for superpositions in the Hilbert space.

Consider the hyperspace vectors $A(t)$, $B(t)$, and $C(t)$ discussed previously, which represent bit strings $\mathrm{A} = 0011, \mathrm{B} = 0101$ and $\mathrm{C} = 0001$. In this example, the XOR operation is implemented between hyperspace vector $A(t)$ and the superposition of $B(t)$ and $C(t)$:

$$A(t)\ \mathrm{XOR}\ [B(t) + C(t)] = A(t)\ \mathrm{XOR}\ B(t) + A(t)\ \mathrm{XOR}\ C(t). \quad (20)$$

Figure 9 demonstrates the XOR operation on a superposition. Likewise, the XNOR operation is implemented between hyperspace vector $A(t)$ and the superposition of $B(t)$ and $C(t)$:

$$A(t)\ \mathrm{XNOR}\ [B(t) + C(t)] = A(t)\ \mathrm{XNOR}\ B(t) + A(t)\ \mathrm{XNOR}\ C(t). \quad (21)$$

Figure 10 illustrates the XNOR operation on a superposition. The outputs of both XOR and XNOR operations were confirmed to be the superpositions representing the sets $\{0110, 0010\}$ and $\{1001,\ 1101\}$ respectively.

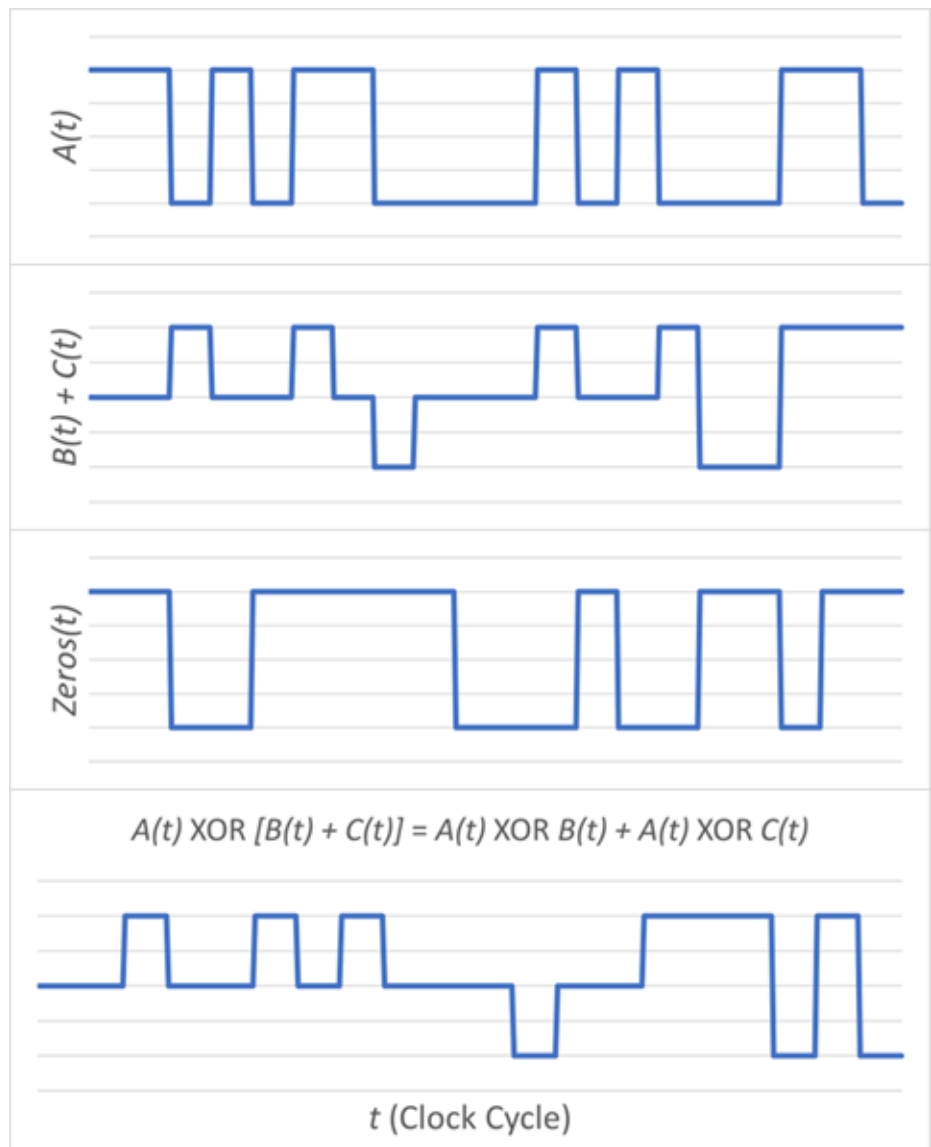

**Figure 9.** The first 20 clock cycles in an $M = 4$ bit INBL system. This illustration shows the implementation of a pairwise XOR operation on a superposition of strings. Where XOR operation is implemented between hyperspace vector $A(t)$ representing $\mathrm{A} = 0011$ and the superposition $B(t) + C(t)$ representing the set $\{\mathrm{B} = 0101, \mathrm{C} = 0001\}$. The output is confirmed to be the superposition representing the set $\{\mathrm{A\ XOR\ B} = 0110,\ \mathrm{A\ XOR\ C} = 0010\}$. Note, the scale of amplitude axis of the superpositions are suppressed.

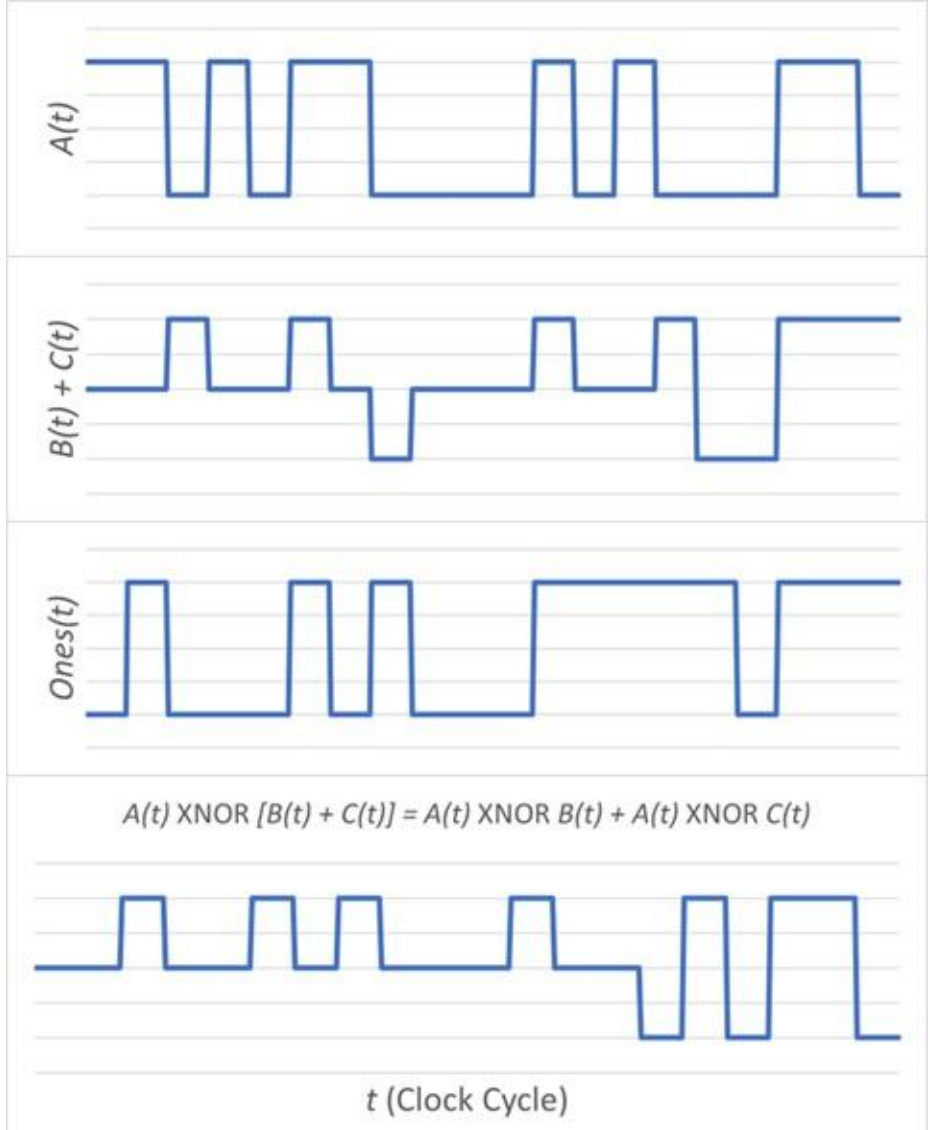

**Figure 10.** The first 20 clock cycles in an $M = 4$ bit INBL system. This illustration shows the implementation of a pairwise XNOR operation on a superposition of strings. The XNOR operation is implemented between hyperspace vector $A(t)$, representing $\mathrm{A} = 0011$, and the superposition $\mathrm{B}(t) + C(t)$, representing the set $\{\mathrm{B} = 0101, \mathrm{C} = 0001\}$. The output is confirmed to be the superposition representing the set $\{\mathrm{A\,XNOR\,B} = 1001, \mathrm{A\,XNOR\,C} = 1101\}$. Note, the scale of amplitude axis of the superpositions are suppressed.

## 3. Conclusions

This work introduces and demonstrates new algorithms for implementing XOR and XNOR operations within the framework of symmetric INBL. Instead of relying on manipulation of the reference signals, the presented approach performs these logic operations directly on strings, hyperspace vectors, or their superpositions. This expands INBL's computational toolkit beyond previously established NOT operations, enabling generalized pairwise and bit-targeted processing without modifications to underlying references. Particularly, the pairwise implementations of the XOR and XNOR operations are a surprising development introduced in this paper.

We reviewed the NOT operation and extended some of its basic concepts for the development of algorithms for the XOR and XNOR operations. Unlike the unary NOT operation, multi-input pairwise operations require additional considerations. Particularly, pairwise operations are only meaningful when both inputs are bit strings with the same length (which includes hyperspace vectors). It is also important to note that the operations presented in this paper can be executed consecutively to accomplish more complex data structure tasks. Future work will explore these applications further.

The proposed algorithms exploit the distributive property of multiplication, allowing operations to propagate across superpositions at constant time complexity, independent of the set size. Simulations using the INBL engine validated the XOR and XNOR operations executed in parallel across sets of numbers (represented by superimpositions). Some limitations were also realized; multiplication-based algorithms cannot distinguish between the cases $j = k = 0$ and $j = k = 1$; see Equations 12 and 13. This limitation poses a challenge for developing the AND and OR operations, which are another consideration for future research.